# DBpia

# Microservices based Framework to Support Interoperable IoT Applications for Enhanced Data Analytics

| | |
|---|---|
| 저자<br>(Authors) | Sajjad Ali, Muhammad Aslam Jarwar, Ilyoung Chong |
| 출처<br>(Source) | 한국통신학회 학술대회논문집 , 2019.1, 636-639(4 pages)<br>Proceedings of Symposium of the Korean Institute of communications and Information Sciences , 2019.1, 636-639(4 pages) |
| 발행처<br>(Publisher) | 한국통신학회<br>Korea Institute Of Communication Sciences |
| URL | http://www.dbpia.co.kr/journal/articleDetail?nodeId=NODE08003467 |
| APA Style | Sajjad Ali, Muhammad Aslam Jarwar, Ilyoung Chong (2019). Microservices based Framework to Support Interoperable IoT Applications for Enhanced Data Analytics. 한국통신학회 학술대회논문집, 636-639 |
| 이용정보<br>(Accessed) | 한국외국어대학교<br>220.67.124.***<br>2019/10/19 15:51 (KST) |

# Microservices based Framework to Support Interoperable IoT Applications for Enhanced Data Analytics

Sajjad Ali, Muhammad Aslam Jarwar and Ilyoung Chong
Department of ICE, Hankuk University of Foreign Studies, Yongin-si, Korea
{sajjad, aslam.jarwar, iychong}@hufs.ac.kr

***Abstract*** — Internet of things is growing with a large number of diverse objects which generate billions of data streams by sensing, actuating and communicating. Management of heterogeneous IoT objects with existing approaches and processing of myriads of data from these objects using monolithic services have become major challenges in developing effective IoT applications. The heterogeneity can be resolved by providing interoperability with semantic virtualization of objects. Moreover, monolithic services can be substituted with modular microservices. This article presents an architecture that enables the development of IoT applications using semantically interoperable microservices and virtual objects. The proposed framework supports analytic features with knowledge-driven and data-driven techniques to provision intelligent services on top of interoperable microservices in Web Objects enabled IoT environment. The knowledge driven aspects are supported with reasoning on semantic ontology models and the data-driven aspects are realized with machine learning pipeline. The development of service functionalities is supported with microservices to enhance modularity and reusability. To evaluate the proposed framework a proof of concept implementation with a use case are discussed.

## I. Introduction

The physical and virtual objects with sensing and actuating capabilities form a vision of the Internet of Things (IoT). The advent of this vision generates tremendous opportunities in different fields. Such as, IoT provides diverse applications in the smart home use case, where home appliances are being integrated with the internet to facilitate ease and comfort in everyday living [1] [2]. Similarly, in the fourth industrial revolution, industries are becoming smarter by integrating their product manufacturing facilities with state of the art IoT technologies in order to monitor and automatically control the production process more rapidly and efficiently than ever before. Smart cities are being envisioned to get enhanced with IoT from traffic management to waste management, from efficiently distributing energy to monitoring public health and safety, every aspect in an urban environment is being designed to be a part of smart IoT environment vision. Many other use cases realize the tremendous power of IoT applications with smart service features [3] [4] [5][6][7].

However, on one hand, it also has been realized that IoT vendors are providing isolated solutions. These solutions although provide some unique functionalities in their own domain, however, they cannot interoperate with other domains. Traditional IoT solutions bring about great challenge of interoperability which hinders the full potential of IoT.

On the other hand, advances in IoT technologies are leading to a growing volume of data which provides a great potential to explore hidden facts about the domain being explored. Advances in machine learning have been improved to analyze the big IoT data more efficiently. Although many analytic solutions are available, still there is a lack of frameworks that support data analytics for heterogeneous data sources.

In view of the above, there are several challenges that need to be solved to support interoperable analytic features on IoT data, these include:

- How heterogeneous objects in different IoT domains can be integrated into a seamless fashion such that uniform service provisioning can be achieved. How to handle diverse data formats of different sensing and actuating devices.
- How to provide interoperable services in a fault tolerant and autonomous plug and play manner so that each service can be replaced in case of failure or in face of changing requirements in a dynamic IoT environment.
- How analytic methods can be incorporated into the design to support efficient analytic operations with reduced complexity and flexibility to work with different data types.

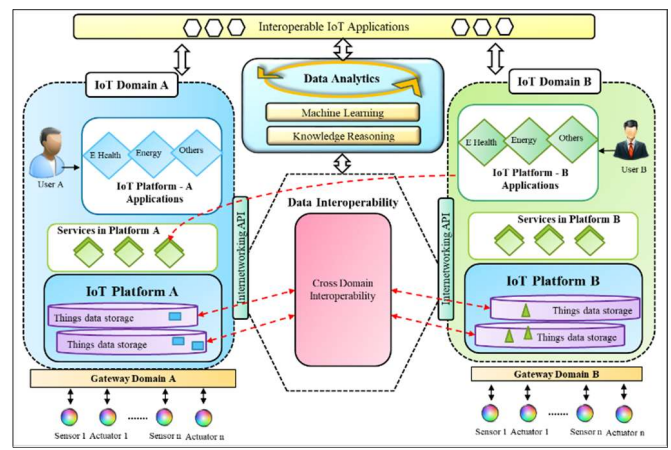

Figure 1. A conceptual model of analytic service provisioning in an interoperable multi-domain IoT environment

To resolve the first challenge we propose using the semantic web technologies integrated with the Web of Objects (WoO) [8] platform. The WoO enables the integration of heterogeneous information sources using semantic ontology models [9]. It supports the virtualization of physical devices and well as other information sources with the Virtual Objects (VOs). The second challenge of fault-tolerant and autonomous services provisioning has been fulfilled through the microservices design architectural pattern [10]. In case of failure or increased service loads, the microservices enable an enhanced scaling of the system dynamically [11]. The third challenge is supported with the integration of multi-domain data with interoperable microservices and on top of this integration intelligent analytics processing layer of knowledge-driven and data-driven microservices have been developed.

To support the above functionalities this article proposes a framework on top of the interoperable microservices which results into uniform processing on heterogeneous IoT data that not only provides analytics on semantic data sources (such as linked data in the form of RDF graphs) through reasoning microservices but also incorporates data-driven pipeline to support machine learning features with microservices.

The interoperable IoT environment involves cross-domain mechanisms [12], where services in one domain can access the features or the data available in another domain with flexibility and transparency. Figure 1 shows a conceptual model of the interoperable environment where cross-domain interoperability infrastructure enables the higher level IoT services such IoT analytics services to provide intelligent service features.

## II. MODELING INTEROPERABLE MICROSERVICES FOR ENHANCED DATA ANALYTICS

In the era of big IoT data, it has become inevitable for IoT platforms to provide tools that can enable powerful analytic features on data to improve and enhance intelligent services. However, as discussed, in multi-domain IoT environment to support efficient analytics on IoT data not only the interoperability should be achieved but such processes should be supported by lightweight and reusable software components [13]. The proposed framework design (illustrated in figure 2) supports the IoT analytics on top of interoperable microservices infrastructure which generates a uniform layer of data objects.

The design is based on a bottom-up style where the data generated from diverse IoT devices are gathered for processing at different levels. It follows the WoO reference platform [14][15] to represent the sensor data. The WoO provides the VO concept which enables a physical object to be digitally represented. Virtualization is one of the major concepts in IoT environments. It provides the digital replacement for the physical objects that can be further utilized in the service workflows at higher levels [16]. VO layer provides the mechanisms to manage virtual objects. VOs are based on the semantic ontology and they follow a well-defined information model which semantically annotates the data coming from the physical or real world objects. The VOs are identified through unique URIs. On top of VO layer, CVOs form a composition of multiple VOs. In other words, one or more VOs are composed in a CVO to serve a service request. It is important to note that CVO not only provides a mashup of multiple VOs but it also defines the set of rules that generate actionable knowledge on VO data. These rules are triggered once the data from the VO satisfy the conditions already defined in the rules of each individual CVO.

In a multi-domain IoT environment, it is necessary to support interoperability in case the data or objects can be reused in different domains. Interoperable microservices layer in our framework besides virtual objects layer incorporates a set of microservices that enable interoperable operations required to mitigate the heterogeneity of different domains.

The annotation microservice enriches the contents from a domain object by linking the information with the semantic ontology. It involves associating extra information with the contents of an object corresponding to an ontology. The annotation of information helps the content to be machine-readable and interpretable in other domains. The query processing microservice handles the queries from the data analytics layer. It checks maintains query logs and executes the query if the entry is available, otherwise, it generates a new query specific to the service requirement. The alignment microservice is invoked in case two objects which need to interoperate in service are using different data models to represent the similar information. In this case, the alignment microservice align the concept from these data models to make them interoperable and usable to handle the service request. The semantic description validator microservice checks the description of annotated contents to verify its correspondence with the semantic ontology concepts and their relationships. Translation microservice translates the data coming from multiple formats to a unified format so that it can be processed by the other services. For example, one task of translation is from relational to RDF translations. Semantic synchronization microservice gets the translated objects from the translation microservices and it synchronizes the objects with respect to other objects that correspond to the same service domain in central objects repository.

The data analytics processing layer provides the functionalities of the core analytic processes using microservices. The analytics processing microservices are categorized using the knowledge-driven and data-driven microservices. The knowledge driven microservices process the semantic data objects and use reasoning engine to infer the knowledge hidden in the semantic objects. The designed set of services are further discussed in section A. In contrast, the data-driven microservices integrate machine learning processes to classify and predict the facts using machine learning library. The developed microservices have been further discussed in section B.

The top layer in the framework is related with the service management functions, which involve system-level functionalities these include service request evaluation and analysis service which processes the user request based on the user history and preference parameters as well as it considers the user access level policy, the lifecycle management process handles the state of different microservices in the system, it takes cares of which microservices are in running state and which are halted due to the service load. The discovery process manages microservices availability by querying the microservice repository database. The microservice template management service maintains the list of templates which are queried in case there are no existing instances in the repository which can satisfy the service request. The microservices repository database provides the entries of all microservices available in the system. Whereas, the orchestration process is responsible for the composition flows of microservices and their configurations.

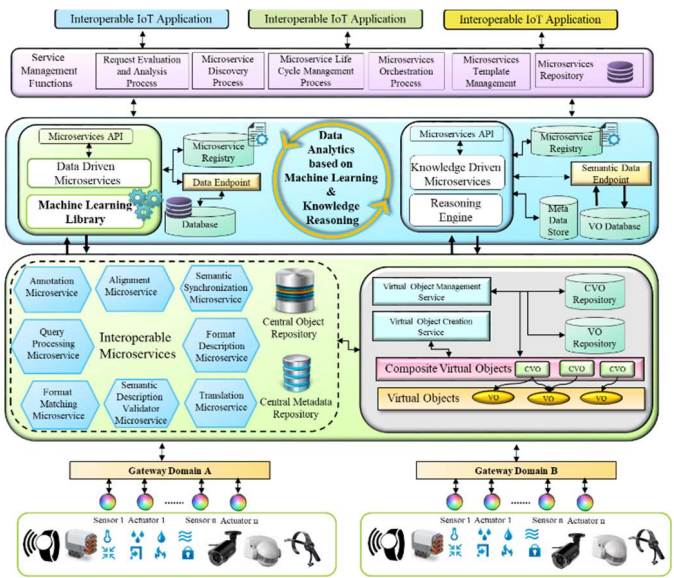

Figure 2. Microservices based framework to support Interoperable IoT applications for enhanced data analytic services

### A. Knowledge-Driven Analytics Model

Based on the proposed framework we have developed microservices to incorporate knowledge driven analytics on IoT data. These microservices utilize real-world knowledge models to provide reasoning based on their underlying structure. Three microservices (shown in figure 3) include the activity reasoning microservice which infers and provides reasoning on the activity ontology model. The location reasoning microservice provides reasoning the location ontology and infers environmental location related facts about a user. The physio status reasoning microservice provides knowledge inference over user physiological data. These knowledge driven microservices work on a well-formed set of rules already defined. Knowledge-driven microservices also access CVO ontologies. As discussed earlier CVO is a composition or a mashup of multiple VOs with the defined rules to extract actionable knowledge.

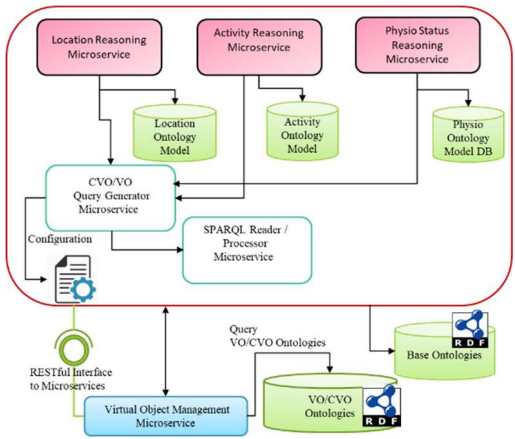

Figure 3. A model of knowledge-driven microservices

Furthermore, the knowledge driven microservices are supported with query generation microservice. This microservice provides the query interface to generate query over VO and CVO objects. The query processing microservice runs the provided query on the virtual objects graphs and returns the semantic results. The virtual object management microservice provides a RESTful interface to support query processing on virtual objects. Also, the Knowledge-driven microservices are supported with the base ontology models in the system these include sensor observation ontology and user model ontology.

### B. Machine Learning based Data Driven Analytics Model

The data-driven analytics microservices (shown in figure 4) are based on the machine learning analysis. Three different microservices are designed to perform the statistical analysis of the IoT data. The location analyzer microservice analysis the user location history and predicts its current location to support service provisioning. The activity analyzer microservice uses the activity history logs of the user based on time series information and predicts the activity to execute user-centered services in IoT environment. The physio status analysis microservice performs prediction on the user current health status and generates a recommendation to the user. These microservices use data analytics manager microservice to read the configuration of multiple models as well they use existing machine learning library to utilize the models trained on the user history data.

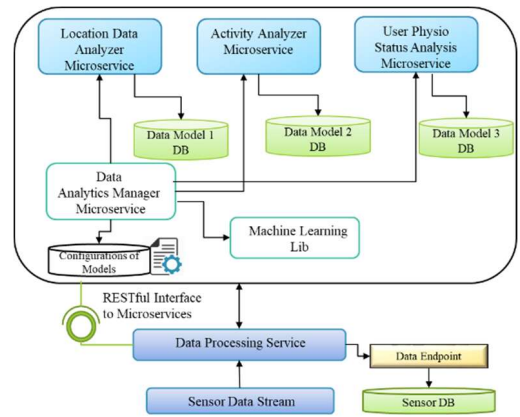

Figure 4. Machine Learning based data-driven analytics model

### III. EVALUATION USE CASE

To evaluate the proposed designed in this paper, a proof of concept prototype has been designed. The system illustrated in the following figure involves there application domains, namely, the smart home domain, the medical facility domain, and the smart office domain. Each domain constitutes an application server which hosts the CVO and VO repository and virtual machines to deploy microservices for the processing of IoT data. All the domains communicate to the central processing hub server which deploys the interoperable microservices and centralized object repository which enables the interoperability of data and services from different domains. The request for any analytic services is handled at the analytics processing server, which hosts request processing and analytic microservice. Based on the service request parameter and checking of the user profile and preference policy, particular microservice is selected for execution. The selected microservice profile is checked at interoperable services hub prior its execution, where if its operations are satisfied from single domain data the service is executed, on the other hand, if multiple domain data is required the central object repository is queried to access the previous service request for extracting the matching object mashups. However in case of no entry is available, the interoperable services hub requests the corresponding domains and generates a mashup to satisfy the requested operations. Moreover, to store VO and CVOs objects Apache Jena framework has been used. RESTful APIs are used for collection of data from multiple IoT domains. To support analytic features Apache Jena inference engine has been used for reasoning and inference. Whereas, for machine learning, WEKA open source library is utilized. The lightweight messaging is implemented with MQTT (RabbitMQ).

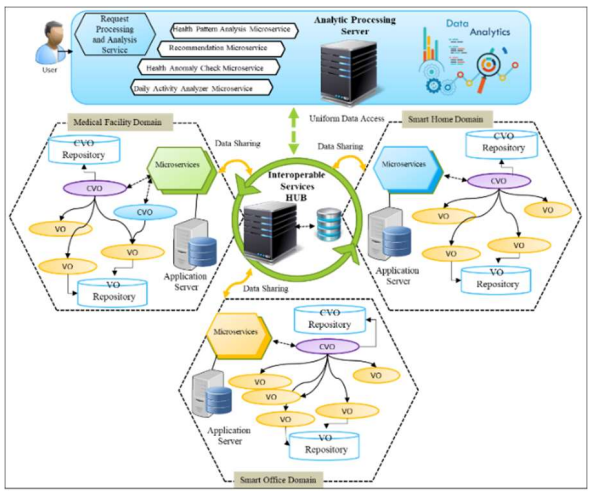

Figure 5.   USE CASE - Multi IoT domain service setup based on microservices

## IV.   CONCLUSION

IoT is growing with an increasing number of IoT objects to provide sophisticated applications that can enhance and automate processes in many real-world domains. However existing IoT platforms have become isolated information silos, where developing applications that involve the data or objects from more than one IoT platform faces a huge problem of interoperability. In this regard, we propose a framework focusing on how interoperable applications can be developed in a multi-domain IoT platform environment. Especially, we have targeted analytics applications supported by interoperable microservices. The contribution of the paper is twofold. First, we propose an interoperability solution for heterogeneous IoT domains and secondly on top of this framework we have designed knowledge-driven and data-driven analytic microservices that provide analytic features on heterogeneous IoT data. Microservices are developed to provide an interoperable medium on diverse data generation and processing components in a uniform manner in the proposed framework.

## ACKNOWLEDGMENT

This research was supported by the MSIT (Ministry of Science and ICT), Korea, under the ITRC (Information Technology Research Center) support program (IITP-2018-0-01396) supervised by the IITP (Institute for Information & communications Technology Promotion)